\documentclass{article}

\usepackage[main, final]{neurips_2026}

\usepackage{amsmath,amsfonts,bm}

\def\eqref#1{equation~\ref{#1}}

\def\1{\bm{1}}

\def\vc{{\bm{c}}}

\def\vh{{\bm{h}}}

\def\vv{{\bm{v}}}

\def\vx{{\bm{x}}}
\def\vy{{\bm{y}}}
\def\vz{{\bm{z}}}

\def\mA{{\bm{A}}}
\def\mB{{\bm{B}}}
\def\mC{{\bm{C}}}
\def\mD{{\bm{D}}}

\DeclareMathAlphabet{\mathsfit}{\encodingdefault}{\sfdefault}{m}{sl}
\SetMathAlphabet{\mathsfit}{bold}{\encodingdefault}{\sfdefault}{bx}{n}

\usepackage{hyperref}       
\usepackage{url}            
\usepackage{booktabs}       
\usepackage{amsfonts}       
\usepackage{nicefrac}       
\usepackage{microtype}      
\usepackage{xcolor}         

\usepackage{microtype}
\usepackage{graphicx}
\usepackage{subcaption}

\usepackage{paralist}
\usepackage{multirow}
\usepackage{colortbl}
\usepackage[algo2e, ruled, lined, linesnumbered, commentsnumbered, longend]{algorithm2e}

\usepackage{hyperref}

\usepackage{amsmath}
\usepackage{amssymb}
\usepackage{mathtools}
\usepackage{amsthm}
\usepackage{comment}

\usepackage[capitalize,noabbrev]{cleveref}

\theoremstyle{plain}
\newtheorem{theorem}{Theorem}[section]
\newtheorem{proposition}[theorem]{Proposition}

\theoremstyle{definition}
\newtheorem{definition}[theorem]{Definition}

\theoremstyle{remark}

\usepackage[textsize=tiny]{todonotes}

\newcommand{\first}[1]{\textbf{#1}}

\newcommand{\proposed}{MFlowAudio}

\definecolor{highlightc}{HTML}{CFF9FA}

\title{MFlowAudio: Efficient Text-to-Audio Synthesis \\ via Mamba-based Stateful Flow Matching}
\author{%
    Hao Dai\textsuperscript{1,2}, Panyu Chen\textsuperscript{1}, Jagmohan Chauhan\textsuperscript{1}\thanks{Corresponding author.}\\
  \textsuperscript{1} Department of Computer Science, University College London\\
  \textsuperscript{2} Department of Engineering, King’s College London\\
  \texttt{\{hao.dai, panyu.chen.25, jagmohan.chauhan\}@ucl.ac.uk} \\
}

\begin{document}

\maketitle

\begin{abstract}

Recent advancements in audio generation have been largely driven by Transformer-based diffusion models. However, these models suffer from  quadratic complexity of self-attention, severely bottlenecking the long-form audio synthesis. To overcome this limitation, we propose MFlowAudio, a novel latent audio generation framework that synergizes the continuous-time dynamics of Flow Matching with a custom-designed TCMamba backbone. TCMamba uses an innovative dual-scan mechanism: a TimeMamba module to capture long-range causal dependencies with linear complexity, and a ChannelMamba module to model cross-channel correlations such as harmonic structures.
Exploiting this structural foundation, we formulate the Stateful Flow Matching (SFM) paradigm. This framework inherently enables chunk-wise training and streaming generation, maintaining an $\mathcal{O}(1)$ caching complexity without incurring extra computational overhead. 
To enable fine-grained controllable synthesis, we devise a novel guidance mechanism that neutralizes vector field collisions precipitated by on-the-fly semantic transitions of prompts.
Comprehensive empirical evaluations confirm that MFlowAudio yields generative fidelity on par with state-of-the-art baselines while establishing significant computational efficiency, achieving above $14.5\%$ acceleration in generation and facilitating perceptually seamless streaming synthesis with a constant latency of $1.8$ seconds.
  Demo:\url{https://huggingface.co/spaces/mflowaudio/MFlowAudio}
\end{abstract}

\section{Introduction}

Recent years have witnessed unprecedented advancements in generative artificial intelligence, revolutionizing domains such as text, image, video, and speech synthesis. A key driver of this progress has been the rapid evolution of diffusion models~\citep{10.5555/3495724.3496298, lu2022dpmsolver,lipman2023flow, tong2024improving}. While these models obtain remarkable achievements in image generation, audio generation has concurrently emerged as a field of increasing research interest and substantial practical impact~\citep{kong2021diffwave, 10.1145/3581783.3612348, huang2023makeanaudio, pmlr-v202-liu23f}.

In practice, many contemporary Text-to-Audio (T2A) models leverage generative architectures originally conceived for the Text-to-Image (T2I) domain, such as U-Net~\citep{ronneberger2015unetconvolutionalnetworksbiomedical} / Transformers~\citep{10.5555/3295222.3295349,dosovitskiy2021an,10377858}-based diffusion backbones.
While early models (e.g., AudioLDM~\citep{pmlr-v202-liu23f, 10.1109/TASLP.2024.3399607}, Make-an-Audio~\citep{10.5555/3618408.3618973, huang2023makeanaudio}, and Tango~\citep{10.1145/3581783.3612348}) achieved success at 16kHz, recent works like Stable-Audio-Open~\citep{evans2024stableaudioopen}, TangoFlux~\citep{hung2024tangofluxsuperfastfaithful}, and SoundCTM~\citep{saito2025soundctm} have shifted to full-band (44.1kHz) synthesis.
As the synthesis of short-form 16kHz audio matures, the research frontier is naturally advancing toward a far more demanding objective: generating extended, continuous full-band audio, such as long-form narratives.

However, modeling long-form, high-fidelity audio presents unique challenges. Continuous audio signals strictly adhere to temporal causality and exhibit notoriously low semantic density per sample. Therefore, preserving acoustic fidelity necessitates extremely high sampling rates, which causes the sequence length to explode as the audio duration scales. This inherent characteristic imposes a high computational burden when attempting to capture long-range dependencies.
Therefore, the Transformer, which is dominant in current generation paradigms, encounters a severe bottleneck, facing two fundamental constraints:
(1) it lacks a strong temporal inductive bias, relying instead on injected positional encodings~\citep{dosovitskiy2021an, 10.1016/j.neucom.2023.127063, kazemnejad2023the} to discern sequence order;
(2) the self-attention mechanism  exhibits computational and memory complexity that scales quadratically (\(\mathcal{O}(L^2)\)) with the sequence length \(L\).

Meanwhile, existing diffusion and flow-matching methods inherently operate on a monolithic, single-pass generation paradigm. This structure exacerbates the memory explosion bottleneck during both training and the generation of long audio sequences, while simultaneously restricting dynamic controllability during generation.
Although autoregressive (AR) models theoretically can generate sequences of arbitrary length, their empirical performance and perceptual fidelity fall short of those of their continuous-time generative counterparts.
This dilemma motivates a critical inquiry: \textit{Is it possible to design an efficient and controllable architecture within the prevailing diffusion/flow-matching paradigm to fulfill the rapidly foreseeable demands of long-form generation?}

To surmount the challenges of computational efficiency and length extrapolation, we propose {\proposed}, a novel T2A architecture.
It integrates the efficient State Space Models (SSMs)~\citep{gu2022efficiently} into a continuous-time generative framework for T2A. Specifically, we introduce the Mamba~\citep{gu2024mamba,mamba2}-based TCMamba block, a core module designed to jointly model the temporal and channel domains, which is optimized via a flow-matching objective.
We demonstrate that this architectural design facilitates chunk-wise training and generation, and explore its capability for real-time streaming generation and dynamic prompt switching, which potentially enhances the practical viability of long-form T2A applications, e.g., interactive agents, real-time game soundscapes and controlled VR audios.
Our core contributions are as follows:

\begin{compactitem}
  \item We propose Time-Channel Mamba, which unifies temporal causality and cross-channel correlation modeling via a novel dual-scan mechanism. It achieves linear complexity with respect to sequence length, eradicating the quadratic memory bottlenecks inherent in Transformer-based continuous-time generative models.

  \item We introduce the Stateful Flow Matching (SFM) algorithm, which natively empowers efficient streaming generation and chunked training with a $\mathcal{O}(1)$ caching complexity and negligible computational overhead. Furthermore, to enable controllable synthesis, we propose a State-Decoupled Conditional Guidance (SDCG) mechanism that mitigates vector field interference during prompt switching.

  \item Extensive experiments demonstrate that {\proposed} achieves generation fidelity highly competitive with state-of-the-art baselines. Furthermore, our framework exhibits   system-level advantages, delivering above $14.5\%$ generation speedup and achieving a perceptually seamless real-time streaming generation with a constant low latency of $1.8$ seconds.
\end{compactitem}

\section{Related Work}

\paragraph{Text-to-Audio Generation.}
Text-to-audio (T2A) generation has rapidly advanced with the adoption of diffusion-based latent models. Early works such as AudioLDM~\citep{pmlr-v202-liu23f} and Make-An-Audio~\citep{10.5555/3618408.3618973} employed U-Net backbones at 16 kHz, demonstrating strong audio quality and text alignment.
Subsequent models, including AudioGen~\citep{kreuk2023audiogen}, Tango~\citep{10.1145/3581783.3612348}, and Stable Audio~\cite{evans2024stableaudioopen}, extended this paradigm to 44.1 kHz (CD quality), significantly improving perceptual fidelity.
To reduce sampling cost, alternatives such as consistency models~\citep{10.5555/3618408.3619743} (e.g., SoundCTM~\citep{saito2025soundctm}) and flow matching approaches (e.g., FreeAudio~\citep{jiangFreeAudioTrainingFreeTiming2025}, TangoFlux~\citep{hung2024tangofluxsuperfastfaithful}) have been introduced, enabling faster generation while maintaining quality. 
However, most models still suffer from a quadratic complexity issue in long contexts due to Transformer-based backbones.

\paragraph{Mamba for Generative Models.}
SSMs and their modern variant, Mamba~\citep{gu2024mamba, mamba2}, have recently demonstrated strong performance in discriminative and autoregressive tasks, as exemplified by VisionMamba~\citep{vim} and Samba~\citep{ren2025samba}.
Building on this momentum, several studies have begun exploring their potential in generative modeling.
For instance, DiMSUM~\citep{phung2024dimsum} integrates Mamba with Transformer in a hybrid architecture to leverage complementary strengths for diffusion-based image generation, while U-Shape Mamba~\citep{Ergasti_2025} adapts SSMs into a U-Net style framework; both designs incorporate specific scanning strategies tailored for image data.
In the audio domain, early efforts such as the SSM-based Sashimi~\citep{pmlr-v162-goel22a} model operate directly on raw waveforms, although its fidelity and text alignment lag behind mainstream T2A models.
Meanwhile, SpeechSSM~\citep{park2025longform}, designed for long-form speech synthesis, demonstrates the potential of SSMs in capturing long-range dependencies.

To the best of our knowledge, {\proposed} is the first framework that applies standalone Mamba to diffusion-based T2A generation, which is specifically designed to natively model the temporal and cross-channel characteristics of audio through its novel dual-path architecture.

\section{Preliminaries}

\paragraph{Diffusion Model and Flow Matching.}
Diffusion models formulate generation as transforming a simple noise distribution into a complex data distribution through a continuous process, typically grounded in the theory of stochastic differential equations (SDEs) or their deterministic counterparts, ordinary differential equations (ODEs). Given a data sample \(\vx_{0} \sim q(\vx_{0})\), the forward process gradually perturbs it into noise by progressively injecting Gaussian noise:
\begin{equation}
q(\vx_{t} | \vx_{t-1} ) =  \mathcal{N} ( \vx_{t} ; \sqrt{1-\beta_t}\vx_{t-1} , \beta_t \mathbf{I})
\end{equation}
The diffusion model learns to reverse the corruption process by means such as parameterizing the denoising process $p_{\theta}(\vx_{t-1} | \vx_t)$  \cite{10.5555/3495724.3496298} or approximating the score function 
\cite{song2020score}.
Flow matching methods \citep{lipman2023flow, albergo2025stochastic} learn a deterministic transformation from noise to data by defining a probability path between a prior distribution $p(z) = \mathcal{N} (0, \mathbf{I})$ and the data distribution $q(\vx_0)$ achieved via the ODE:
\begin{equation}
 \frac{d\vx_t}{dt} = v_{\theta}(\vx_t, t)
\end{equation}
where $v_{\theta}(\vx_t, t)$ is a time-dependent vector field learned to regress the conditional flow $u_t(\vx_t|\vz)$ from noise $\vz$ to data sample $\vx_0$. 
The training objective minimizes the difference between the model's velocity and the target vector field under the conditional probability path:
\begin{equation}
 \mathcal{L}_{FM} = \mathbb{E}_{t, p(\vx_t|\vz), p(\vz)} [ || v_{\theta}(\vx_t, t) - u_t(\vx_t | \vz) ||^2 ]
\end{equation}
Once trained, the model $v_{\theta}$ defines a deterministic mapping from the prior to the data distribution.
To generate a sample, one first draws $\vz \sim p(\vz)$ and then solves the probability flow ODE from $t=0$ to $t=1$ using any numerical ODE solver (e.g., Euler or Runge-Kutta methods).

Diffusion and flow matching models employ a parallel decoding paradigm, where the entire sequence is decoded over multiple steps.
Let $L$ denote the sequence length and NFE (Number of Function Evaluations) represent the total number of sampling iterations.
The computational cost can then be approximated as $c(v_{\theta}(L)) \times \text{NFE}$, where $c(v_{\theta}(L))$ is the cost per iteration. Accordingly, efficiency can be improved along two primary directions: reducing NFE or optimizing $c(v_{\theta}(L))$.

\paragraph{Transformer v.s. Mamba.}

The transformer backbone is ubiquitous in generative modeling. A common pipeline is to patchify the input into patches, project each patch to a fixed-dimension embedding, add positional encodings, and model global interactions via self-attention over the resulting token sequence.
While effective at capturing long-range spatial dependencies, self-attention requires pairwise interactions among all tokens and thus incurs $\mathcal{O}(L^2)$ computation and memory for a sequence of length L. Moreover, because self-attention is permutation-invariant, it depends on injected positional bias (e.g., RoPE) to encode order, which hampers extrapolation to longer contexts.

To enable more efficient audio generation, we investigate RNN-like inductive biases, with Mamba as a promising candidate. Mamba builds upon a continuous state space representation, which is discretized for practical implementation. Its core SSM follows a recurrent formulation:
\begin{equation}
 \vh_{t} = \mA \vh_{t-1} + \mB \vx_{t}; \quad \vy_{t} = \mC \vh_{t} + \mD \vx_{t} 
\end{equation}
while a novel selection mechanism makes the transition and input matrices ($\mB$ and $\mC$) input-dependent, enabling context-dependent representations. Furthermore, a hardware-aware parallel algorithm ensures efficient training despite the recurrent nature of the architecture.
These properties make Mamba well-suited for long-form audio generation: it captures long-range dependencies with linear complexity $\mathcal{O}(L)$, while preserving temporal coherence across extended contexts.

A limitation of Mamba is its design for one-dimensional sequential data, while the dominant practice encodes audio as 2D time–frequency spectrograms, which are more compact and preserve essential acoustic structure.
Mamba's native 1D recurrence, however, is not well matched to spectral patterns and cross-channel correlations.
To bridge this gap and leverage Mamba’s efficiency for long-range dependencies, we introduce a novel Mamba-based block tailored to operate on 2D spectrogram.

\section{\proposed}

 In this section, we will introduce the core design of {\proposed} (Mamba-based Flow-Matching Text-to-Audio).

\subsection{Time-Channel Mamba Block}

State-of-the-art generative models for audio typically encode signals into a two-dimensional latent representation of shape $(L, C)$, where $L$ denotes length and $C$ denotes channels.
For instance, Stable Audio Open~\citep{evans2024stableaudioopen}  uses a VAE to encode audio into a 64-channel 2D latent. 
Unlike image latents, the temporal axis of audio latents is separable: a latent of shape $(L, C)$ can be partitioned into multiple contiguous chunks along the length dimension: 
\begin{equation}
  \begin{aligned}
    & (L,C) = \{(L^{(1)},C), (L^{(2)},C), \dots (L^{(k)},C)\} \\
    & L = L^{(1)} + L^{(2)} + \dots + L^{(k)}
  \end{aligned}
\end{equation}
each of which can be independently decoded into a coherent audio clip. 
This motivates us to design a time-latent oriented Mamba block that respects temporal separability while maintaining efficiency.

\begin{figure}[t]
\setlength{\abovecaptionskip}{-1pt}
\setlength{\belowcaptionskip}{-10pt}
\begin{center}
  \includegraphics[width=.95\linewidth]{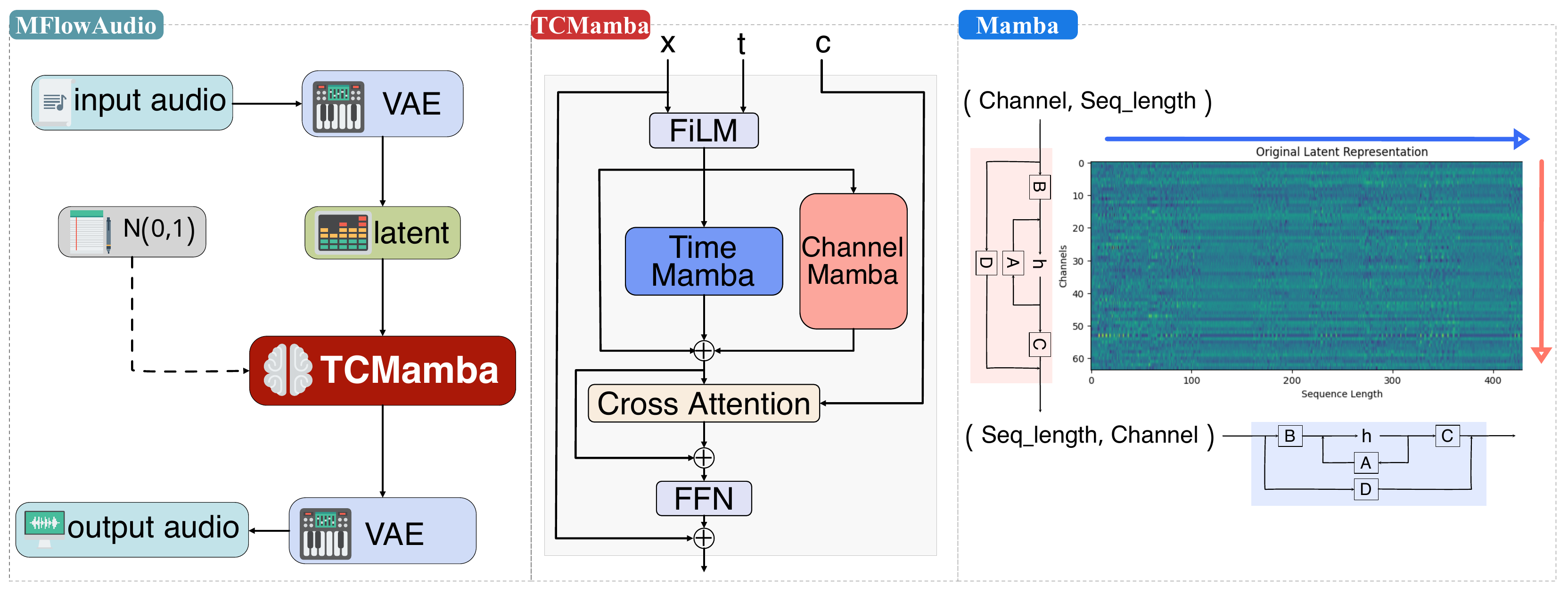}
\end{center}
  \caption{The key components of {\proposed}: (left) the workflow of {\proposed}, audio is encoded and decoded into latent using VAE and vocoder; (middle) the architecture of the TCMamba block; and (right) the dual-scanning mechanism of Mamba. }
\label{fig:tfmaudio}
\vspace{-1mm}
\end{figure}

\paragraph{Time Mamba.}
Intuitively, audio can be treated as sequential data and modeled directly using Mamba. In our proposed TimeMamba block, the audio latent representation is processed as a sequence of shape $(1, C)$ along the temporal dimension. As illustrated in \textit{Figure~\ref{fig:tfmaudio}(right)}, the model scans the sequence from left to right, updating a hidden state $\vh_t$ at each step to predict the subsequent representation. This design offers several key advantages:
\begin{compactitem}[-]
\item Inherent Temporal Modeling: The RNN-like recurrent state update naturally preserves chronological relationships and causal structure, which is fundamental to audio signals.

\item Physically-Grounded Dynamics: SSMs are inherently designed to describe dynamical systems. Since audio is itself a physical phenomenon generated by a dynamical process, this formulation offers a more natural inductive bias for audio generation.

\item Linear Computational Efficiency: By replacing the quadratic complexity of self-attention, TimeMamba achieves $\mathcal{O}(L)$ complexity with respect to sequence length.
\end{compactitem}

Despite TimeMamba’s strong capability in temporal modeling, its key shortcoming is the limited ability to capture coherent cross-channel structures, such as harmonic resonances across channels. 

\paragraph{Channel Mamba.}
To overcome the limitation in cross-channel resonance modeling, we introduce ChannelMamba, a variant that reinterprets the latent as a sequence along the channel axis.
ChannelMamba scans along the channel, which means the input shape is $(1, L)$, thereby capturing cross-channel correlations and harmonic structures. 
Since scanning across $C$ requires a fixed length, which limits its ability to process audio of arbitrary length, we exploit the temporal separability of audio and instead scan shorter subsequences $(1, L^{(i)})$, enabling processing variable-length input while preserving local harmonic relationships.

The dual-scan mechanism is illustrated in \textit{Figure~\ref{fig:tfmaudio}(right)}: TimeMamba operates along the temporal dimension while ChannelMamba scans along the channel axis.
By integrating FiLM and Cross-Attention for condition injection, we formulate the proposed TCMamba block, as depicted in \textit{Figure~\ref{fig:tfmaudio}(middle)}. Comprehensive architectural specifications are detailed in Appendix~\ref{appendix:model}.

\paragraph{Computational Complexity Analysis.}
For TimeMamba and ChannelMamba, both components build upon  SSM, whose complexity scales linearly with sequence length. For an input latent $\vx \in \mathbb{R}^{L \times C}$, the complexity of applying SSM updates is $\mathcal{O}(L \cdot C)$.
For FiLM Modulation, it introduces minimal overhead, requiring only element-wise affine transformations. Therefore, the FiLM operation has complexity $\mathcal{O}(L \cdot C)$, maintaining the linear scaling characteristics of the overall architecture. 
For Cross-Attention: While cross-attention between the latent sequence $(L, C)$ and conditioning embeddings $(S, C)$ has complexity $\mathcal{O}(L \cdot S \cdot C)$, where $S$ is the length of the conditioning sequence, in practice $S$ is fixed and $\ll L$ for typical audio generation.

The overall complexity of TCMamba can be summarized as:  $\mathcal{O}(L \cdot C + L \cdot S \cdot C) \approx \mathcal{O}(L)$ assuming $C$ and $S$ are constants or significantly smaller than $L$. This linear scaling enables efficient processing of long audio sequences while maintaining strong modeling capabilities. 

\paragraph{Flow Matching Training.}

We adopt the Optimal Transport Flow Matching (OT-FM) framework~\citep{tong2024improving} to train our generative model. OT-FM provides a simulation-free training objective, avoiding costly stochastic trajectory sampling, and instead relies on analytically defined optimal transport paths for efficient and stable learning.
Formally, let $\vx_0 \sim \mathcal{N}(0, \mathbf{I})$ denote the initial noise distribution and $\vx_1 \sim q(\vx)$ be the target audio latent representation.
For each timestep $t \in [0, 1]$, we define the linear interpolation path:
\begin{equation}
\vx_t = t\, \vx_1 + (1-t)\, \vx_0
\end{equation}
Our parameterized vector field $v_{\theta}(\vx_t,t, \vc)$, conditioned on timestep $t$ and semantic signal $\vc$, is trained to approximate this velocity field $(\vx_1 - \vx_0)$. The Flow Matching loss is therefore:
\begin{equation}
  \begin{aligned}
    \mathcal{L}_{FM} = \mathbb{E}_{\vx_0 \sim \mathcal{N}(0,\mathbf{I}),\, \vx_1 \sim q(\vx),\, t \sim \mathcal{U}[0,1]} 
    \Big[ \big\| v_{\theta}(\vx_t, t, \vc) & - (\vx_1 - \vx_0) \big\|^2 \Big]
  \end{aligned}
\end{equation}

\subsection{Stateful Flow-Matching}

Given the recurrent design of Mamba, we design a novel Stateful Flow Matching (SFM) framework. Standard flow matching relies on a monolithic generation process, which imposes memory constraints when scaling to long sequences. Furthermore, extending it to generate consecutive temporal chunks typically necessitates intricate context-conditioning mechanisms to maintain temporal coherence.
By exploiting the state-space representations, SFM circumvents these limitations by introducing an additional state cache. It achieves chunk-wise training and generation at near-zero additional computational cost with a constant cache, while maintaining the generation quality of the original one-time generation.

\begin{figure}[t]
\setlength{\abovecaptionskip}{-2pt}
\setlength{\belowcaptionskip}{-2pt}
\begin{center}
  \includegraphics[width=.95\linewidth]{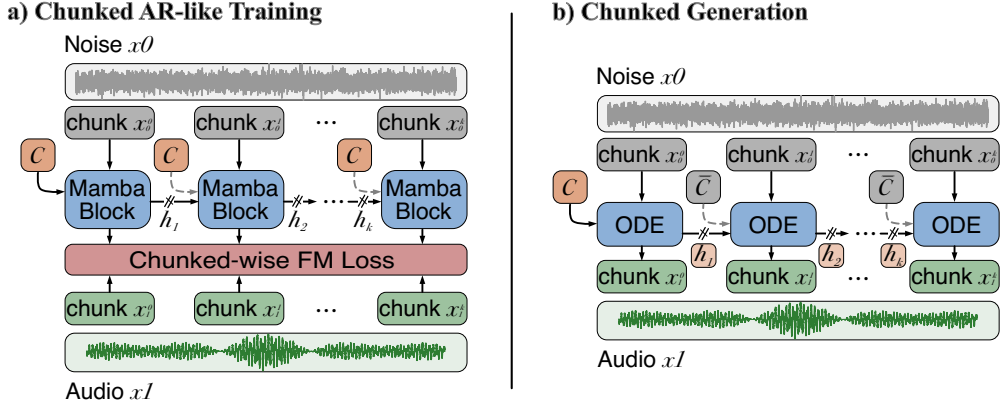}
\end{center}
  \caption{Beyond conventional monolithic training and generation, {\proposed} also facilitates (a) chunked AR-style training and (b) chunked generation to support streaming.}
\label{fig:chunked}
\vspace{-1mm}
\end{figure}

\subsubsection{Chunked Training and Generation}
To this end, we augment TCMamba block with an explicit state output, ensuring that each forward pass returns the terminal hidden state $\vh_i$. Given that the Mamba is a sequential scanning structure, the state evolution along the temporal axis dictates that training or generation for any given chunk can resume from the final state of the preceding chunk. Consequently, by caching this terminal state, TCMamba supports chunk-wise inference (as shown in Figure~\ref{fig:chunked}), remaining equivalent to the conventional single-pass paradigm (see Appendix~\ref{appendix:chunked} for detailed theoretical analysis):
\begin{equation}
\begin{aligned}
  & \vec{v}_{i}, \vh_{i+1} = v_{\theta}(\vc; \vh_i) \\
  & \vec{v} = \{ v_{\theta}(\vc; \vh_0), v_{\theta}(\vc; \vh_1), \cdots ,v_{\theta}(\vc; \vh_i)\}
\end{aligned}
\end{equation}

\paragraph{$\mathcal{O}(1)$-Memory Temporal Propagation.} The computational overhead introduced by this chunk-wise processing paradigm is negligible. Firstly, it incurs absolutely zero additional computational burden. Secondly, because the mechanism only requires caching the terminal state, the auxiliary memory cost for state propagation remains $\mathcal{O}(1)$ w.r.t. sequence length $L$. 
In contrast, the benefit in memory efficiency is substantial. 
By processing data strictly within a fixed chunk size of $L^{(c)}$ (where $L^{(c)} < L$ and is  a constant), the overall spatial complexity is reduced to $\mathcal{O}(L^{(c)})$, thereby enabling the training and generation of long sequences under constrained memory budgets.

\paragraph{Streaming Generation and Time-to-First Latency.} 
Beyond memory efficiency, the chunked formulation inherently enables real-time streaming generation. Because audio is consumed sequentially, playback commences immediately upon the synthesis of the initial chunk. Provided the per-chunk inference time remains strictly less than its acoustic duration (i.e., a Real-Time Factor $< 1$), {\proposed} sustains a seamless pipeline capable of generating arbitrary-length audio with deterministic latency. This paradigm minimizes user-perceived delay, bounding it entirely by the Time-To-First-Chunk (TTFC). Furthermore, from a deployment perspective, the temporal amortization of computation profoundly alleviates serving bottlenecks, significantly maximizing server capacity and throughput for concurrent requests. See Appendix~\ref{appendix:streaming} for details of implementation.

\subsubsection{Dynamic and Controllable Conditional Generation}

We further extend continuous generation by dynamically switching prompts at chunk boundaries. This mechanism brings benefits as it enables fine-grained temporal control (e.g., triggering vocal onset at 3.7s) and flexible compositional prompting.  While some existing works attempt to achieve this via densely annotated datasets and bespoke architectures, our approach achieves temporal alignment through the intervention during chunked generation.
\begin{figure}[htb]
\vspace{-1mm}
\setlength{\abovecaptionskip}{-1pt}
\setlength{\belowcaptionskip}{-1pt}
\begin{center}
  \includegraphics[width=.55\linewidth]{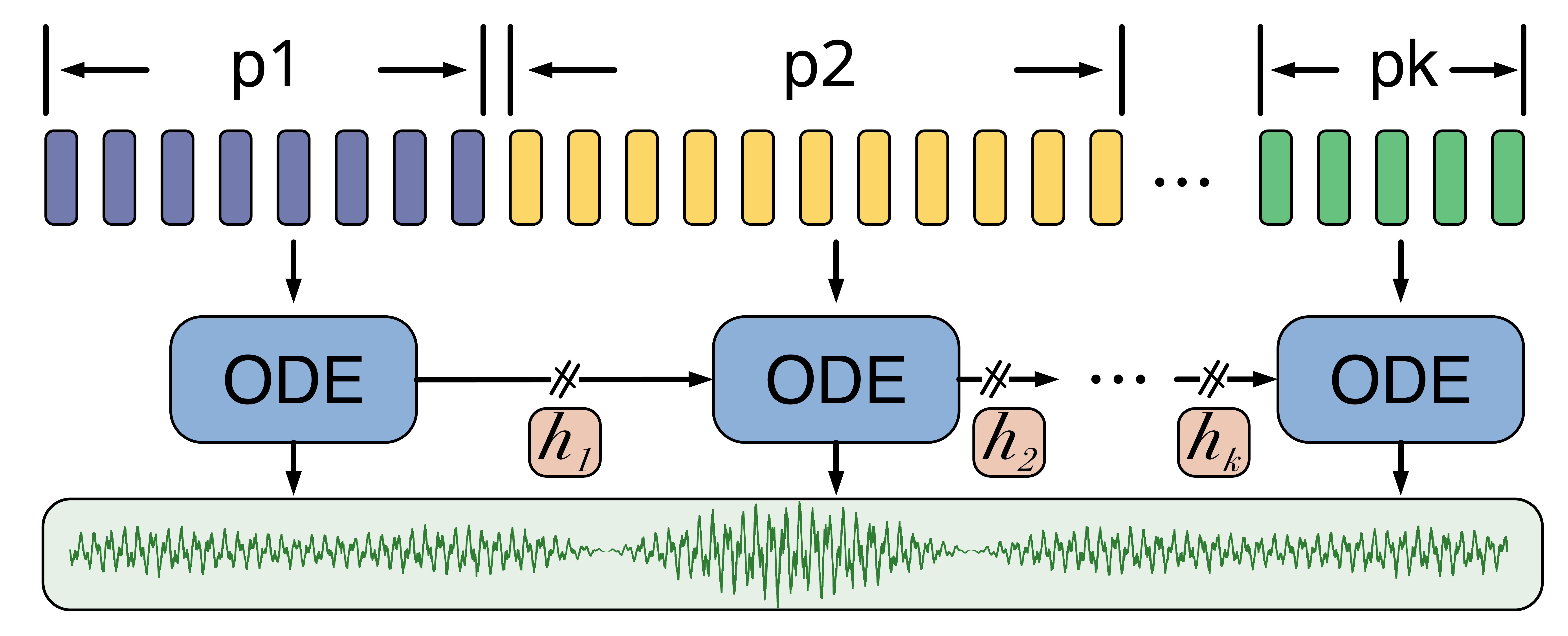}
\end{center}
  \caption{Switching prompts across variable-sized chunks to control the exact timing of events.}
\label{fig:dyna-gen}
\vspace{-1mm}
\end{figure}

\paragraph{Dual-Driven Dynamics: Balancing State Inertia and Condition Injection.}
A critical issue of prompt switching is maintaining acoustic consistency across chunk boundaries, such as spatial reverberation, amplitude, and phase coherence. Due to the standard FM driving by prompt in a memoryless manner, generating consecutive chunks with distinct prompts inevitably introduces artifacts, e.g, audible clicks and phase discontinuities.
In contrast, SFM is a dual-driven dynamic system, its vector field is governed by two distinct velocities: the current text condition and the historical state. When the prompt is null or transitions, SFM naturally relies on the "historical inertia" ($\vec{v}_{\text{state}}$) preserved within the states to perpetuate the prior semantic and acoustic properties:
\begin{equation}
\vec{v}_{\text{state}} = \vv_{\theta}(\emptyset; \vh_i); \quad \vec{v} = \vec{v}_{\text{uncond}} + \omega_{\text{cfg}} * ( \vec{v}_{\text{state}} - \vec{v}_{\text{uncond}} )
\end{equation}
This quasi-Markovian "inertia" guides the generation trajectory to glide smoothly along the data manifold. By functioning as a continuous regularizer across chunk boundaries, this state-driven momentum ensures acoustic coherence and spontaneously generates a natural transition zone.

\paragraph{State-Decoupled Conditional Guidance.}
When a semantically contradictory prompt is injected, the SFM confronts a severe control contention. In such circumstances, the model's vector field is simultaneously pulled by two antagonistic driving forces. This tension invokes vector field collision along the ODE integration trajectory. The resulting instability within the vector field dynamics induces destructive signal interference, leading to the representation collapse of either the historical state or the newly injected prompt, thereby degrading the generation quality.

To resolve this dilemma, we introduce a strategy that prioritizes the fidelity of the newly injected prompt while maximally retaining compatible historical state information. 
To achieve this, it is essential to extract the state-aware semantic vector, $\vec{p}$, from the composite vector field, which is distinct from conventional stateless $v_{\text{prompt}} = \vv_{\theta}(\vc, \emptyset)$. Specifically, we leverage the unconditional stateful vector field, $\vec{v}_{\text{state}}$, to explicitly decouple the prompt's pure contribution from the overall conditional vector field, $\vec{v}_{\text{cond}}$. Furthermore, by stripping away unconditional interference, we confine the guidance within the disentangled semantic subspace:
\begin{equation}
\begin{aligned}
  \vec{c} &\triangleq \vec{v}_{\text{cond}} - \vec{v}_{\text{uncond}}, \quad \vec{s} \triangleq \vec{v}_{\text{state}} - \vec{v}_{\text{uncond}}, \quad \vec{p} = \vec{c} - \vec{s}  \\
\end{aligned}
\end{equation}
Ideally, the semantic direction $\vec{p}$ and the state direction $\vec{s}$ should remain orthogonal to prevent mutual interference. Therefore, to mitigate the vector field collision, we propose an Orthogonal Score Projection mechanism to extract and preserve the component of $\vec{p}^{\perp}$ that is orthogonal to $\vec{s}$, while truncating the component of $\vec{p}^{\parallel}$ that is parallel to and co-directional (reverse to preserve more history) with $\vec{s}$. This approach reinforces the impact of the new prompt within the pure semantic subspace while simultaneously preserving the underlying acoustic skeleton, thereby minimizing adverse state interference. The final rectified vector field is formulated as follows:
\begin{equation}
\label{eq:sdcg}
\begin{aligned}
  \vec{v}^{\prime} &= \vec{v}_{\text{uncond}} + \omega_{\text{cfg}}\!\left[\, \vec{c} - \max\!\left(0,\; \tfrac{\langle \vec{c},\, \vec{s}\rangle}{\lVert \vec{s}\rVert^{2}} - \delta\right) \vec{s} \,\right].
\end{aligned}
\end{equation}
This State-Decoupled Conditional Guidance (SDCG) mechanism amplifies the guidance strength of the prompt with only an additional unconditional-stateful forward pass at the boundaries. It also introduces an adjustable scalar parameter $\delta \in [0, 1]$ to attenuate the historical state intensity. This $\delta$ serves as a versatile condition during generation, allowing for flexible, fine-grained control over the state-to-prompt influence ratio. See Appendix~\ref{appendix:sdcg} for more details.

\section{Experiments}
Following established practices in latent audio models, we adopt the VAE from Stable Audio Open as our audio encoder and utilize the T5~\citep{2020t5} model as our text prompt encoder. The backbone of {\proposed} is constructed by stacking multiple TCMamba blocks, with detailed architectural parameters provided in the Appendix~\ref{appendix:model}.

\subsection{Experimental Setup}

\textbf{Datasets and Training Setup.}
We trained our model using the AudioCaps~\citep{kim-NAACL-HLT-2019} and WavCaps~\citep{mei2023wavcaps} (AudioSet-SL subset only) training sets. All audio clips were resampled to 44.1 kHz. For testing, we evaluated our model on the AudioCaps full test set ($4352$ samples), which was also resampled to 44.1 kHz.
We trained our model using $8$ NVIDIA H100 GPUs with a batch size of 64 per GPU for $100$ epochs.
See Appendix~\ref{appendix:training} for the details of training and generation.

\textbf{Baselines.}
We compared our model against three state-of-the-art full band (44.1 khz) text-to-audio generation models: AudioLDM2-48K~\citep{10.1109/TASLP.2024.3399607}, Stable Audio Open~\citep{evans2024stableaudioopen}, and TangoFlux~\citep{hung2024tangofluxsuperfastfaithful}.
Stable Audio Open, TangoFlux, and {\proposed} all share the same VAE architecture (Oobleck) and a T5-family text encoder.
In addition, we also compared with an autoregressive 16khz model AudioGen~\citep{kreuk2023audiogen}.
For each baseline, we utilized the officially released pre-trained checkpoints and adhered to their recommended configurations to report their best performance.

\textbf{Evaluation Metrics.}
To evaluate the generated audios, we employ six established metrics covering both intrinsic audio quality and text-audio alignment: CLAP~\citep{laionclap2023} Score (text alignment), PANNs~\citep{10.1109/TASLP.2020.3030497}-based Inception Score (quality/diversity), VGGish~\citep{7952132}-based and OpenL3~\citep{8682475}-based Frechet Audio Distance (FAD$_{v}$ and FAD$_{openl3}$)~\citep{kilgour2019frechetaudiodistancemetric} / Frechet Distance (FD) (quality/distributional match), and KL divergence (distributional match).
To ensure a fair comparison and facilitate reproducibility, all evaluation metrics are computed using the publicly available audioldm-eval toolkit~\citep{pmlr-v202-liu23f}. 

\subsection{Results}

\paragraph{Main Results.}
To validate the efficacy of our proposed architecture, we conduct a comprehensive comparison of generation quality. It is worth noting that TangoFlux, in addition to being trained on the audio dataset (TangoFlux-base), underwent fine-tuning using an auxiliary CRPO dataset. For comparison, we fine-tuned {\proposed} utilizing a training set synthesised by TangoFlux. As presented in Table~\ref{tab:main}, under a near equivalent parameter budget, {\proposed} demonstrates superior performance across primary acoustic quality metrics (FD, FAD$_v$, FAD$_{openl3}$ and KL divergence). While slightly trailing TangoFlux in the CLAP and IS score, our model achieves an overall performance highly competitive with the current SOTA. These results suggest that TCMamba possesses audio processing capabilities comparable to Transformer-based architectures, underscoring its practical viability.
We also conducted an evaluation on Clotho dataset, see Appendix~\ref{appendix:ood}.

\begin{table}[t]
  \caption{Performance comparison across baselines on AudioCaps testset.}
\label{tab:main}
\begin{center}
\resizebox{0.95\textwidth}{!}
{\begin{tabular}{@{}lllllllllll@{}}
    \toprule
\multicolumn{1}{l}{\bf Duration} & \multicolumn{1}{l}{\bf Model}  &\multicolumn{1}{c}{\bf{Parameters}} &\multicolumn{1}{c}{\bf FD${\downarrow}$} &\multicolumn{1}{c}{\bf KL${\downarrow}$} &\multicolumn{1}{c}{\bf IS${\uparrow}$}   &\multicolumn{1}{c}{\bf FAD$_{v}$${\downarrow}$} &\multicolumn{1}{c}{\bf CLAP${\uparrow}$}  &\multicolumn{1}{c}{\bf FAD$_{openl3}$${\downarrow}$}
\\ \midrule 

    \multirow{7}{*}{10s}
& AudioGen (16K) &  1.5B &38.7259 &2.5699 & 6.6716 &6.6353 &0.1950 & 192.4442 \\ 
& AudioLDM2-48K & 1.7B &36.2054& 2.0261 & 5.442 & 5.4774& 0.3827& 148.1751 \\
& Stable Audio Open & 1B & 32.0744& 2.052 &10.1462 & 3.7894 &0.3442 & 109.0337 \\
  & TangoFlux-base &  516M & 16.9351 & 1.1724 & \first{12.0486} & 3.2514& \first{0.4722}&  113.6228 \\ 
  &\cellcolor{highlightc!60}{\proposed}-base & \cellcolor{highlightc!60}581M & \cellcolor{highlightc!60}\first{10.9178}& \cellcolor{highlightc!60}\first{1.1588} & \cellcolor{highlightc!60}10.8935 & \cellcolor{highlightc!60} \first{2.9612} & \cellcolor{highlightc!60} 0.4655 & \cellcolor{highlightc!60} \first{95.6402} \\ \cmidrule{2-9}
  &TangoFlux &  516M & 15.8429& 1.1164& \first{12.5489}& \first{2.1963}& \first{0.4990}& 102.8903 \\
  & \cellcolor{highlightc!60}{\proposed} & \cellcolor{highlightc!60}581M & \cellcolor{highlightc!60}\first{11.8569} & \cellcolor{highlightc!60} \first{1.1158} & \cellcolor{highlightc!60}11.6865& \cellcolor{highlightc!60}2.5248 & \cellcolor{highlightc!60}0.4957& \cellcolor{highlightc!60}\first{96.2363}  \\

  \midrule
    \multirow{6}{*}{30s} 
& AudioLDM2-48K & 1.7B & 52.5380 &2.3984 & 5.1880 &10.8777 &0.2750 & 175.1954 \\
& Stable Audio Open& 1B & 28.1238 & 1.9539 & 10.1849&4.3256& 0.3619& 108.4538 \\
& TangoFlux-base & 516M  &24.6576& 1.2224 & 9.0941 & 4.9399 & 0.4173 & 123.9447 \\
  & \cellcolor{highlightc!60}{\proposed}-base &\cellcolor{highlightc!60} 581M &\cellcolor{highlightc!60}\first{17.1580}& \cellcolor{highlightc!60}\first{1.1491} & \cellcolor{highlightc!60}\first{10.6577} & \cellcolor{highlightc!60}\first{3.6937} & \cellcolor{highlightc!60}\first{0.4333} & \cellcolor{highlightc!60}\first{103.0790} \\ \cmidrule{2-9}
  & TangoFlux & 516M &20.1872& \first{1.1189} & 10.7179 & 2.9555 & \first{0.4541} & 102.2544 \\
  & \cellcolor{highlightc!60}{\proposed} & \cellcolor{highlightc!60}581M &\cellcolor{highlightc!60}\first{17.5387}& \cellcolor{highlightc!60}1.1411 & \cellcolor{highlightc!60}\first{11.3356} & \cellcolor{highlightc!60}\first{2.8981} & \cellcolor{highlightc!60}0.4448 & \cellcolor{highlightc!60}\first{94.4700} \\

\bottomrule
\end{tabular}}
\end{center}
\vspace{-1mm}
\end{table}

\paragraph{Generation Latency and Real-Time Streaming.} To rigorously evaluate computational efficiency, we benchmarked the generation latency of all models under identical hardware conditions (a single RTX-5080 16GB GPU) using default configurations. For robustness, all reported latencies are averaged over $20$ independent generation runs. As in Table~\ref{tab:latency}, {\proposed} demonstrates highest efficiency, achieving a speedup of $81.8\%$ for 10-second and $14.5\%$ for 30-second over baselines.
Besides computational complexity, another reason affecting latency is that both StableAudio and TangoFlux generate a fixed-length latent (their maximum duration) and trim the output afterward, i.e., they always run the same amount of compute during denoising regardless of the target length, while {\proposed} adapts its compute to the target length and generates audio.

\begin{table}[htb]
  \caption{Generation latency comparison (s)}
  \label{tab:latency}
\begin{center}
\resizebox{0.8\textwidth}{!}
{
  \begin{tabular}{@{}ccccc@{}}
    \toprule
  \multicolumn{1}{l}{\bf Duration}  & \multicolumn{1}{l}{\bf AudioLDM2-48K} & \multicolumn{1}{l}{\bf Stable-Audio-Open} & \multicolumn{1}{l}{\bf TangoFlux} & \multicolumn{1}{l}{\bf {\proposed}} 
\\ \midrule 
    10s & 22.23$\pm$3.1 s & 5.23$\pm$13 ms & 3.51$\pm$3 ms & \textbf{1.93}$\pm$13 ms \\
    20s & 32.17$\pm$4.3 s & 5.31$\pm$21 ms & 3.54$\pm$12 ms & \textbf{2.17}$\pm$17 ms \\
    30s & 97.82$\pm$7.5 s & 5.45$\pm$25 ms & 3.55$\pm$5 ms & \textbf{3.10}$\pm$18 ms  \\
\bottomrule
\end{tabular}
  }
\end{center}
  \vspace{-1mm}
\end{table}

Furthermore, we conducted a Time-To-First-Chunk (TTFC) evaluation to validate the model's streaming capabilities. As shown in Table~\ref{tab:ttfc}, with a chunk size set to 64 (corresponding to approximately $3$ seconds of audio), the initial decoding time is $~1.8$ seconds. 
Because this generation time is strictly shorter than the chunk's playback duration (i.e., Real-Time Factor $< 1$), perceptually seamless streaming generation is fully realizable. Regardless of the total target length, the user experiences a fixed wait time of $1.8$ seconds before playback commences. Additionally, when increasing the batch size, decoding remains highly stable, yielding a maximum latency reduction of approximately $63.9\%$.

\begin{table}[htb]
\centering
  \caption{One-time Latency v.s. Time-to-First-Chunk  (s)}
\label{tab:ttfc}
\resizebox{0.7\textwidth}{!}
{
\begin{tabular}{@{}cccccc@{}}
\toprule
  \multicolumn{1}{l}{\bf Duration} & \multicolumn{1}{l}{\bf BatchSize} & \multicolumn{1}{l}{\bf One-time} & \multicolumn{1}{l}{\bf TTFC} & \multicolumn{1}{l}{\bf Speedup} & \multicolumn{1}{l}{\bf Reduction} \\
\midrule
  10.0 & 1 & 1.9${\pm}$19 ms & 1.82${\pm}$9 ms & 1.1x & +5.8\%    \\
10.0 & 2 & 2.0${\pm}$13 ms & 1.85${\pm}$21 ms & 1.1x & +10.6\%   \\
20.0 & 1 & 2.1${\pm}$26 ms & 1.81${\pm}$17 ms & 1.2x & +16.8\%   \\
20.0 & 2 & 3.5${\pm}$32 ms & 1.85${\pm}$31 ms & 1.9x & +47.4\%   \\
30.0 & 1 & 3.1${\pm}$18 ms & 1.80${\pm}$29 ms & 1.7x & +41.7\%   \\
30.0 & 2 & 5.1${\pm}$57 ms & 1.85${\pm}$37 ms & 2.8x & +63.9\%   \\
\bottomrule
\end{tabular}
  }
\vspace{-1mm}
\end{table}

\paragraph{Dynamic Injection and Boundary Smoothness Evaluation.}
To evaluate the model's capacity for dynamic prompt injection, we curated a subset of $1267$ prompts from the standard benchmark that exhibit explicit temporal transitions (e.g., "followed by," "then," "after"). Each prompt was decomposed into a sequence of two to three distinct sub-prompts, which were uniformly anchored to predetermined onset timestamps at 0s, 3s, and 6s.
Furthermore, to objectively quantify the acoustic continuity at the chunk boundaries (i.e., prompt switching points), we introduce two specialized signal processing metrics: Boundary Maximum Spectral Flux (BMaxSF) and Background Median Spectral Flux (BgMedSF). Building upon these, we formulate the Boundary-to-Background Ratio ($BBR = \frac{BMaxSF}{BgMedSF}$). A lower BBR indicates that the spectral variation at the transition boundary closely matches the natural variance of the background context, thereby serving as a robust indicator of seamless, artifact-free transitions (see Appendix~\ref{appendix:transition} for detailed definition and analysis.).

\begin{table}[htb]
  \caption{Performance comparison of dynamic conditional injection}
\label{tab:dyna}
\begin{center}
\resizebox{0.9\textwidth}{!}
{\begin{tabular}{@{}lllllllll@{}}
    \toprule
    \multicolumn{1}{l}{\bf Model} &\multicolumn{1}{c}{\bf FD${\downarrow}$} &\multicolumn{1}{c}{\bf KL${\downarrow}$} &\multicolumn{1}{c}{\bf IS${\uparrow}$}   &\multicolumn{1}{c}{\bf FAD$_{v}$${\downarrow}$} &\multicolumn{1}{c}{\bf CLAP${\uparrow}$}  &\multicolumn{1}{c}{\bf FAD$_{openl3}$${\downarrow}$}  &\multicolumn{1}{c}{\bf{BBR}${\downarrow}$}
\\ \midrule 

  TangoFlux (OneTime)& 21.2885& 1.1381 & 9.9338 & 2.1743 & 0.5208& 110.7247&  - \\
  TangoFlux (Dyna-P)&22.6873& 1.2788 & 6.2233 & 3.9077& 0.4191& 114.2842  & 11.6443  \\ \midrule{}
  {\proposed} (OneTime) &16.3070& 1.1416& 9.2517 & 2.2228& 0.5186& 96.8583 & -  \\
  {\proposed} (Dyna-P)& 15.8575 & 1.2190 & 7.5717 & 2.8448& 0.5087 & 101.2825 & \textbf{4.3287} \\ 

\bottomrule
\end{tabular}}
\end{center}
\vspace{-1mm}
\end{table}

The results are presented in Table~\ref{tab:dyna}. To establish an intuitive baseline, we included standard one-time generation as a reference. The data reveals that dynamic prompt injection incurs minimal degradation in overall audio fidelity; specifically, there is virtually no performance drop in CLAP and FD scores, confirming that the generated audio remains highly viable for practical applications. 
Furthermore, to benchmark transition smoothness, we evaluated the spectral flux metrics against TangoFlux (despite its lack of native support for dynamic prompting). As demonstrated, {\proposed} exhibits a  low BBR,  indicating significantly smoother and more natural transitions across semantic boundaries.

\begin{table}[htb]
  \caption{Ablation experiments of SDCG}
\label{tab:abl}
\small
\begin{center}
\resizebox{0.9\textwidth}{!}
{\begin{tabular}{@{}lllllllll@{}}
    \toprule
    \multicolumn{1}{l}{\bf Model} &\multicolumn{1}{c}{\bf FD${\downarrow}$} &\multicolumn{1}{c}{\bf KL${\downarrow}$} &\multicolumn{1}{c}{\bf IS${\uparrow}$}   &\multicolumn{1}{c}{\bf FAD$_{v}$${\downarrow}$} &\multicolumn{1}{c}{\bf CLAP${\uparrow}$}  &\multicolumn{1}{c}{\bf FAD$_{openl3}$${\downarrow}$}  &\multicolumn{1}{c}{\bf{BBR}${\downarrow}$}
\\ \midrule 

  w/o state &19.3239& 1.3015& 6.8395 & 3.5030& 0.4925& 99.4748& 7.4001 \\
non-state prompt& 16.9351 & 1.2767 & 7.2420 & 3.1169 & 0.4996& \textbf{99.2386} & 5.4953 \\
  vanilla &16.0940& 1.2536& \textbf{7.6370} & 2.9659& 0.5060& 102.8763 & \first{4.2633} \\
w/ naive sub &16.0056& 1.2494& 7.6201 & 2.8947& 0.5075& 102.3913 & 4.3578 \\
w/ sdcg& \textbf{15.8575} & \textbf{1.2190} & 7.5717 & \textbf{2.8448}& \textbf{0.5087} & 101.2825 & 4.3287 \\

\bottomrule
\end{tabular}}
\end{center}
\vspace{-1mm}
\end{table}

\paragraph{Ablation Studies for SDCG.} To validate the efficacy of proposed SDCG, we conducted a comprehensive set of ablation experiments. 
As detailed in Table~\ref{tab:abl}, we evaluated SDCG against four alternative configurations: 1) w/o state: omitting state propagation entirely and relying instead on overlap-add techniques at boundaries; 2) non-state prompt: driving the ODE via a naive linear combination of stateless conditional field $\vec{v}_{\text{prompt}}$ and $\vec{v}_{\text{state}}$; 3) vanilla: utilizing the unadjusted conditional vector field $\vec{v}_{\text{cond}}$; and 4) naive subtraction: applying a direct difference $\vec{v}_{\text{cond}} - \vec{v}_{\text{state}}$.

The results demonstrate that SDCG is best on the fidelity/alignment metrics (FD, KL, $FAD_v$, CLAP) while trading off on IS, $FAD_{openl3}$, and BBR.
This outcome is highly intuitive from a dynamic perspective: intensifying the prompt guidance inherently induces more decisive semantic shifts, whereas the vanilla model relies predominantly on state inertia, preserving acoustic smoothness at the cost of prompt adherence. This phenomenon highlights a trade-off between precise semantic control and boundary naturalness.

\section{Conclusion and Discussion}

We introduced {\proposed}, a latent flow matching framework for text-to-audio generation that marries the efficiency of State Space Models with robust cross-channel modeling. By unifying TimeMamba and ChannelMamba via a dual-scan architecture, we realized a Stateful Flow Matching paradigm that circumvents both the quadratic complexity and the monolithic generation constraints of traditional Transformer-based diffusion models. Our approach delivers high-fidelity audio synthesis on standard benchmarks with high efficiency. 
In addition, its native support for chunk-wise streaming minimizes user-perceived latency, bringing the potential for real-time interactive systems and accommodating on-the-fly semantic interventions.

Currently, the dynamic control mechanism operates in a zero-shot manner, it relies purely on inference-time intervention rather than explicitly learning the underlying dynamics of acoustic transitions from data.
The state guidance achieves transitions by mixing audio in different proportions, rather than by genuinely understanding how different elements should be composed.
The model would benefit from further optimization on rich, sequentially annotated multi-prompt datasets.
In addition, BBR mainly reflects changes at spectral boundaries and cannot capture other perceptual aspects, such as artifacts.
And the subject evaluation should include more extensive A/B tests, providing a more comprehensive perceptual assessment of single-pass versus chunked generation, the SDCG ablations, and dynamic switching.
Addressing these limitations will be the primary focus of our future research to unlock more complex controllable generation applications.

\section*{Acknowledgments}

This work was supported by the Engineering and Physical Sciences Research Council (EPSRC) grant, \textit{MultiTasking and Continual Learning for Audio Sensing Tasks on
Resource-Constrained Platforms [EP/X01200X/1]}, and NVIDIA Academic Grant, \textit{CoVLA - Cloud-Edge Collaborative Data Generation Framework for VLA Learning}.

\newpage

\bibliography{main}
\bibliographystyle{plainnat}

\newpage

\appendix

\section{Appendix}

\subsection{Model Configurations}
\label{appendix:model}

{\proposed} follows a latent diffusion paradigm with the following components:

\paragraph{Audio VAE.}
We employ the VAE from Stable-Audio-Open as our latent audio encoder and vocoder.
This VAE features a 64-channel bottleneck and achieves a compression factor of 2048.
For a 44.1 kHz audio signal, the encoder produces a latent representation of shape $(L, 64)$, where the sequence length $L$ is calculated as $L = \text{duration} \times 44100 / 2048 $.

\paragraph{Text Encoder.}
We utilize the T5-base model as our text encoder.
For each input prompt, the encoder produces a fixed-dimensional representation with a shape of $(128, 768)$, corresponding to 128 token embeddings each of dimensionality 768.

\paragraph{Time and Channel Mamba.}
Our TimeMamba module is implemented based on the official Mamba2~\citep{mamba2} source code (\url{https://github.com/state-spaces/mamba.git}), with a model dimension of 768 and a state dimension of 128. Similarly, ChannelMamba follows the same implementation but with a model dimension of 64 while maintaining the same state dimension of 128.

\paragraph{Conditional Generation.}
Conditioning signals guide the model toward semantically aligned outputs. Following the standard formulation of conditional diffusion models, the network input is represented as a triplet $(\vx_t,t,\vc)$, where $\vx_t$ is the noisy latent at timestep $t$, $t$ is the diffusion timestep embedding, and $\vc$ denotes the conditioning signal (e.g., text prompt embeddings).

For temporal conditioning, we adopt Feature-wise Linear Modulation (FiLM)~\citep{10.5555/3504035.3504518}, in which timestep $t$ is projected via a learned embedding function.

\emph{FiLM.} The timestep embedding is generated using a sinusoidal positional encoding, which is then projected through two linear layers with a GELU activation in between to produce the parameters $(\gamma_t, \beta_t)$ with shape (1, 768), which are then applied within each TCMamba block to modulate activations:

\begin{equation}
\text{FiLM}(\vx_t,t) = \gamma_t \odot \vx_t + \beta_t
\end{equation}

For semantic conditioning, we incorporate cross-attention to align the noisy latent representation with the conditioning signal.

\emph{Cross-Attention.} The cross-attention mechanism employs $16$ attention heads, and the total model dimension is 768. The key and value projections are derived from the T5 text embeddings, while the query projection is obtained from the latent representation. It is defined as:
\begin{equation}
\text{CrossAttn}(Q,K,V) = \text{Softmax}(\frac{QK^T}{\sqrt{d}})V 
\end{equation}
where $Q = W_Q \cdot \vx_t, K = W_K \cdot \vc, V = W_V \cdot \vc$, and $d$ is the feature dimension for scaling. This design enables the model to dynamically attend to the most semantically relevant components of c, ensuring faithful alignment between conditioning inputs and generated audio.

\paragraph{Overall Architecture.}
The overall architecture begins with an initial linear projection layer that maps the 64-channel latent input to a 768-dimensional feature space. This is followed by a stack of 48 identical TCMamba modules. Each TCMamba module contains a TimeMamba block and a ChannelMamba block, augmented with FiLM and cross-attention. The network concludes with a final linear projection layer that maps the 768-dimensional features back to the original 64-channel latent space.

\subsection{Training and Sampling Details}
\label{appendix:training}

The training procedure, outlined in \textit{Algorithm~\ref{alg:training}}, minimizes the loss function depicted in \textit{Figure~\ref{fig:train}}. 
The AdamW optimizer was employed with an initial learning rate of \(1 \times 10^{-4}\), decayed using a cosine annealing scheduler over two decay cycles.
During generation, as detailed in \textit{Algorithm~\ref{alg:sampling}}, we employ an Euler solver with a fixed step size of 0.04, and the CFG scale $\omega_\text{cfg}$ was set to $2.5$.

\begin{algorithm2e}[htb]
  \caption{Stateful Flow Matching Training}
  \label{alg:training}
    \DontPrintSemicolon
    \SetKwInOut{KwIn}{Input}
    \SetKwInOut{KwOut}{Output}

    \KwIn{datasets \((q(\vx), \vc), v_{\theta}, t_{\text{scheduler}}, \epsilon=0.1\)}

    \For {epoch in epochs}{
     \tcp{Sample data}
     Sample \(\vx_1 \sim q(\vx), \vx_0 \sim \mathcal{N}(0, \mathbf{I}), t \sim t_{\text{scheduler}}\)\;
    \If{\(random() < \epsilon\)}{
      \(\vc \leftarrow \emptyset\)
    } \Else{
      \(\vc \leftarrow \text{Encoder}(\vc)\)
    }

     \tcp{Interpolate}
     \(\vx_t \leftarrow t \vx_1 + (1-t) \vx_0\)\;

\BlankLine
\BlankLine

    \tcp{full sequence training}
     \tcp{Predict velocity}
     \(\hat{v_{\theta}} \leftarrow v_{\theta}(\vx_t, t, \vc)\)\;
     \tcp{Compute loss}
     \(\mathcal{L}_{FM} \leftarrow ||\hat{v_{\theta}} - (\vx_1 - \vx_0)||^2\)\;

     \BlankLine
\BlankLine

     \tcp{chunked-wise training}
     \For{each chunk \(i\)}{
       \(\vx_{0}^{i} \leftarrow \vx_0[i]\) ,\, \(\vx_{1}^{i} \leftarrow \vx_1[i]\) \;
        \(\vx_t^i \leftarrow t \vx_1^i + (1-t) \vx_0^i\)\;
        \tcp{Predict velocity}
        \(\hat{v}_{\theta}^i, h^i \leftarrow v_{\theta}(\vx_t^i, t, \vc, h^{i-1})\)\;
      \tcp{Compute loss}
     \(\mathcal{L}_{FM}^{i} \leftarrow ||\hat{v}_{\theta}^i - (\vx_1^i - \vx_0^i)||^2\)\;
     }
     \tcp{Chunked-wise FM loss}
    \(\mathcal{L}_{FM} \leftarrow \sum_i \mathcal{L}_{FM}^{i}\)

\BlankLine
\BlankLine

     Update \(v_{\theta}\) using AdamW optimizer\;
    }
\end{algorithm2e}

\begin{algorithm2e}[htb]
  \caption{{\proposed} Generation}
  \label{alg:sampling}
    \DontPrintSemicolon
    \SetKwInOut{KwIn}{Input}
    \SetKwInOut{KwOut}{Output}

    \KwIn{\(v_{\theta}, \omega, \delta , \vc,\) step-size=0.04}

    \(\vx_t \sim \mathcal{N}(0, \mathbf{I})\) \;

      \BlankLine
      \tcp{One-time Generation}

    \For {t in [0, 1] with step-size}{
      \(\vec{v}_{cond} \leftarrow v_{\theta}(\vx_t, t, c)\)\;
      \(\vec{v}_{uncond} \leftarrow v_{\theta}(\vx_t, t, \emptyset)\)\;
      \(\vec{v} \leftarrow \vec{v}_{uncond} + \omega ( \vec{v}_{cond} - \vec{v}_{uncond} ) \)\;
      \(\vx_t \leftarrow \vx_t + \text{step-size} \cdot \vec{v}\)\;

    }

      \BlankLine
      \BlankLine
      \tcp{Chunked generation with SDCG}
      \For{each chunk \(L_{c}\)}{
        \(\vx_t^c \leftarrow \vx_t[c] \)\;
        \For {t in [0, 1] with step-size}{
          \(\vec{v}_{cond}^c, h^c \leftarrow v_{\theta}(\vx_t^c, t, c, h^{c-1})\)\;
          \(\vec{v}_{uncond}^c, h^{c}_u \leftarrow v_{\theta}(\vx_t^c, t, \emptyset, h^{c-1}_u)\)\;
          \eIf{\(\vc^{c} \ne \vc^{c-1}\)}{
            \(\vec{v}_{state}^c \leftarrow \vec{v}_{uncond}^c \)\;
            \(\vec{v}_{uncond}^c, h^{c}_u \leftarrow v_{\theta}(\vx_t^c, t, \emptyset, h^{0}_u) \)\;
            \(\vec{v}^c \leftarrow \text{SDCG}(\omega, \vec{v}_{cond}^c, \vec{v}_{state}^c, \vec{v}_{uncond}^c)\) \;
          }{
            \(\vec{v}^c \leftarrow \vec{v}_{uncond}^c + \omega ( \vec{v}_{cond}^c - \vec{v}_{uncond}^c ) \)\;
        }
        \(\vx_t^c \leftarrow \vx_t^c + \text{step-size} \cdot \vec{v}^c\)\;
      }
    }
\end{algorithm2e}

\subsection{Subjective Evaluation}

We conducted a subjective evaluation with $25$ participants using a blind listening test to assess randomly selected audio samples. Participants rated the samples across two dimensions: Text Relevance (REL) and Overall Quality (OVL), both on a 5-point Likert scale. As shown in Table~\ref{mos}, our approach slightly outperforms TangoFlux regarding overall quality, highlighting the effectiveness of the MFlowAudio framework in enhancing generative fidelity.

We further conducted a subjective evaluation of the audio generated at prompt switches, focusing on artifacts, clicks/pops, and transition naturalness. Three metrics were used: Boundary Artifact Score (BAS): Perceived severity of audible discontinuities at chunk boundaries; Click/Pop Occurrence (CPO): Presence of transient clicks or pops in the generated audio; and Transition Naturalness (TN): Smoothness and plausibility of transitions when the text prompt is switched. 
The results (Table~\ref{natu-eval}) show that our method produces more natural transitions with fewer clicks and pops.

\begin{figure}[htb]
\begin{center}
  \includegraphics[width=1.0\linewidth]{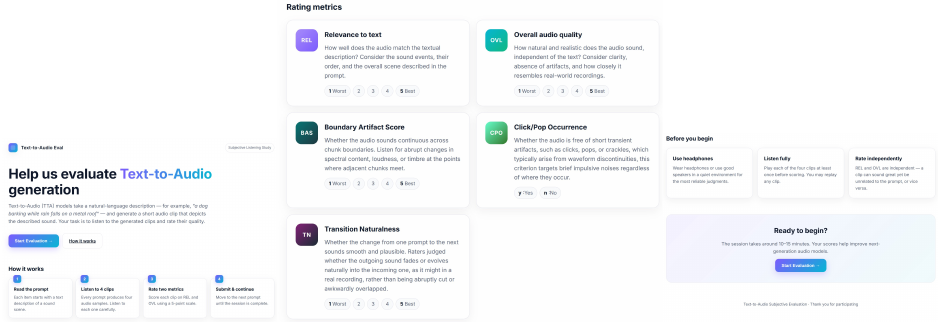}
\end{center}
\caption{evaluation platform for subjective evaluations.}
\label{fig:subeval}
\end{figure}

\begin{table}[htb]
  \caption{Subjective evaluations}
  \label{mos}
\small
\begin{center}
  \begin{tabular}{@{}lll@{}}
    \toprule
    \multicolumn{1}{l}{\bf Model} & \multicolumn{1}{l}{\bf REL${\uparrow}$}  &\multicolumn{1}{c}{\bf OVL${\uparrow}$}
\\ \midrule 
    AudioLDM2-48K & 2.61 & 2.57  \\
    Stable-Audio-Open & 2.82 & 3.28  \\
    TangoFlux & 4.15 & 3.52  \\
    {\proposed} & 4.09 & 3.59  \\
\bottomrule
\end{tabular}
\end{center}
\end{table}

\begin{table}[htb]
  \vspace{-3mm}
  \caption{Naturalness evaluations with dynamic prompting}
  \label{natu-eval}
\small
\begin{center}
  \begin{tabular}{@{}llll@{}}
    \toprule
    \multicolumn{1}{l}{\bf Model} & \multicolumn{1}{l}{\bf BAS${\uparrow}$}  &\multicolumn{1}{c}{\bf CPO${\downarrow}$} &\multicolumn{1}{c}{\bf TN${\uparrow}$}
\\ \midrule 
    TangoFlux (Dyna-P) & 4.08 & 0.76 & 2.04  \\
    {\proposed} (Dyna-P) & 4.12 & 0.02 & 4.18 \\
\bottomrule
\end{tabular}
\end{center}
\end{table}

\subsection{Equivalence of Chunked and One-Time Integration}
\label{appendix:chunked}

In Flow Matching models, the generation trajectory is solved by integrating the Ordinary Differential Equation (ODE). With the assumption of a constant semantic condition (i.e., the text prompt $c$ remains invariant), we establish that the state-passing chunked integration is mathematically strictly equivalent to the monolithic one-time global integration. This equivalence provides a theoretical guarantee for our model's capacity to generate continuous, seamless audio sequences.

\begin{definition}[Global and Streaming ODE Systems]
  Let the global target audio sequence be denoted as $\vx = [\vx^{(1)}, \vx^{(2)}, \dots, \vx^{(K)}]$, where $\vx^{(i)}$ represents the $i$-th chunk. Assuming the existence of an ideal causal global vector field $\vec{v}_{global}$, its corresponding global ODE system is formulated as:
  $$ \frac{d\vx_t}{dt} = \vec{v}_{global}(\vx_t, t), \quad X_0 \sim \mathcal{N}(0, I) $$
In our proposed Stateful Flow Matching (SFM) architecture, historical context is continuously propagated through the hidden state $h_{i-1}(t)$. Consequently, the piecewise streaming ODE system is defined as:
$$ \frac{d\vx_t^{(i)}}{dt} = \vv_\theta(\vx_t^{(i)}, t, h_{i-1}(t)), \quad \text{where } h_i(t) = f(h_{i-1}(t), \vx_t^{(i)}) $$
where $t \in [0, 1]$ denotes the integration time mapping from the noise distribution to the data distribution.
\end{definition}

Building upon the definition of the streaming ODE system, we can establish the explicit relationship between the global vector field and its piecewise counterpart.

\begin{proposition}[Vector Field Isomorphism]
  \label{prop:vfi}
Assume the model architecture satisfies strict temporal causality, and the hidden state mapping $\mathcal{H}$ serves as a sufficient summary of the historical information, namely $h_{i-1}(t) \equiv \mathcal{H}(\vx_t^{(1)}, \dots, \vx_t^{(i-1)})$. With these conditions, the spatial concatenation of the streaming local vector fields is strictly isomorphic to the global causal vector field.
\end{proposition}

\begin{proof}
  Given that the global vector field $\vec{v}_{global}$ is strictly causal, its component corresponding to the $i$-th chunk depends exclusively on the current chunk and its preceding history:
  $$ \vec{v}_{global}^{(i)}(\vx_t, t) = \hat{v}(\vx_t^{(i)}, t, \vx_t^{(1:i-1)}) $$
  In our proposed SFM architecture, the hidden state $h_{i-1}(t)$ recursively encodes the entire causal history $\vx_t^{(1:i-1)}$. Consequently, there exists a parameterized local vector field $\vv_\theta$ such that:
  $$ \vv_\theta(\vx_t^{(i)}, t, h_{i-1}(t)) \equiv \vec{v}_{global}^{(i)}(\vx_t, t) \quad \forall i \in \{1, \dots, K\}, \forall t \in [0, 1] $$
  This implies that at any given integration timestep $t$, the spatial concatenation of the piecewise local dynamics driven by the state-passing mechanism is mathematically equivalent to the holistic global dynamics.
\end{proof}

Given this proposition, it can be deduced that, with the same initial noise, the integration trajectories of one-time and chunked generation are equivalent.

\begin{theorem}[Trajectory Equivalence]
  \label{thm:equiv}
  Given an identical initial Gaussian noise distribution $\vx_0 \sim \mathcal{N}(0, I)$ and assuming constant semantic conditioning, the concatenated solution trajectory $\tilde{\vx}_t = [\tilde{\vx}_t^{(1)}, \dots, \tilde{\vx}_t^{(K)}]$ obtained via piecewise integration in SFM is strictly equivalent to the global trajectory $\vv_t$ solved by the monolithic ODE system. That is, $\tilde{\vx}_t = \vx_t, \forall t \in [0, 1]$.
\end{theorem}
\begin{proof}
  According to the Picard-Lindelöf theorem from fundamental ODE theory (which establishes the existence and uniqueness of solutions to initial value problems): 

  Assuming the vector field $\vec{v}_{global}(\vx, t)$ is locally Lipschitz continuous with respect to $\vx$ across the state space, then for any given initial condition $\vx_0$, there exists a unique global solution trajectory $\vx_t$.

  Now, consider the piecewise integration process of SFM. As established in Proposition~\ref{prop:vfi}, each sub-ODE $\frac{d\vx_t^{(i)}}{dt} = \vv_\theta(\vx_t^{(i)}, t, h_{i-1}(t))$ fundamentally constitutes an exact subsystem of the global ODE framework $\frac{d\vx_t}{dt} = \vec{v}_{global}(\vx_t, t)$.

 Because the initial condition $\tilde{\vx}_0 = \vx_0$ is strictly identical and the driving vector fields are equivalent everywhere, the concatenated trajectory $\tilde{\vx}_t$ derived from piecewise integration is necessarily a valid solution to the aforementioned global initial value problem.

Furthermore, since the Picard-Lindelöf theorem guarantees the absolute uniqueness of this global solution, it inevitably follows that:
$$ \tilde{\vx}_t \equiv \vx_t, \quad \forall t \in [0, 1] $$
\end{proof}

\subsection{Implementation of State Caching and Chunked Generation}
\label{appendix:streaming}

\paragraph{Per Step State Caching.}
Building upon the established Theorem~\ref{thm:equiv}, we can conclude that as long as the underlying model maintains strict causality, caching the hidden state $h_t$ is sufficient to achieve chunk-wise integration that is mathematically identical to monolithic global integration. However, because the numerical ODE solver in Flow Matching operates over multiple discrete timesteps, strict equivalence must be meticulously preserved at every single intermediate step. 

To ensure this, our framework explicitly caches and propagates $h_t$ at each discrete integration step along the trajectory. The total number of integration steps is typically defined by a fixed constant, commonly referred to as the Number of Function Evaluations (NFE). In our default generation configuration, the NFE is set to 25, which corresponds to a uniform integration step size of 0.04.

\paragraph{Causal Caching in TCMamba.}
Our overall framework effectively satisfies the causality prerequisite outlined in Proposition~\ref{prop:vfi} along the critical temporal dimension.
Specifically, our model employs a dual-scan mechanism comprising two functionally distinct modules: a TimeMamba and a ChannelMamba. While the ChannelMamba scans orthogonally along the channel axis and does not adhere to temporal causal constraints, the TimeMamba strictly enforces unidirectional causality along the time axis. Since the theoretical requirement for seamless chunk-wise generation relies exclusively on temporal continuity, we streamline our state-passing mechanism by only caching and propagating the hidden states of the TimeMamba, bypassing the ChannelMamba.

The cache holds the per-layer TimeMamba SSM states across the NFE integration steps, and the two CFG branches, i.e., its size is $(2 \times \text{NFE} \times \text{per-layer states})$, a fixed constant in absolute size. It does not grow with audio duration (unlike a Transformer KV-cache, which grows $\mathcal{O}(L)$). 

The state passed between chunks is the per-block Mamba2 SSM state, shape
$(B,24,64,128)$, it has no length dimension. A matched-parameter Transformer
backbone (self-attention, temporal width 800) has no such compressed state and must
retain every chunk's keys/values.

\begin{table}[htb]
  \caption{Cache size comparison between MFlowAudio and Transformer-based diffusion models}
\label{tab:cache}
\small
\begin{center}
\resizebox{0.7\textwidth}{!}
{\begin{tabular}{@{}lll@{}}
    \toprule
    \multicolumn{1}{l}{\bf $L$ (dur.)} & \multicolumn{1}{l}{\bf MFlowAudio SSM state / fwd} &\multicolumn{1}{c}{\bf Transformer KV cache / fwd} \\ \midrule 
   128 (~6 s) & \bf 18.0 MB & 18.8 MB  \\
   256 (~12 s) & \bf 18.0 MB & 37.5 MB \\
   512 (~24 s) & \bf 18.0 MB & 75.0 MB \\
   scaling & \bf constant, $\mathcal{O}(1)$ & $4\times$, $\mathcal{O}(L)$ \\

\bottomrule
\end{tabular}}
\end{center}
\end{table}

As shown in Table~\ref{tab:cache}, the SSM cache is identical at 6 s, 12 s and 24 s (18 MB fp16) and adds zero extra
FLOPs; the full streaming-generation cache is
$18\,\text{MB}\times\text{NFE}\times2\,(\text{CFG})\approx 900$ MB, still independent
of duration. The Transformer KV cache instead grows linearly ($4\times$ over
$6\!\rightarrow\!24$ s), and being bidirectional it cannot stream causally at all.

\paragraph{Implementation of Streaming Generation.}

For the backbone, the cached SSM state is a sufficient statistic only for the SSM recurrence. Mamba2's causal conv1d retains an additional finite state (its preceding frames) that the cache omits. Because this second assumption is only approximately met, we obtain approximate equivalence, which we mitigate with a 3-frame warmup and quantify as a bounded 2–6\% residual.

As per the VAE, since the Stable Audio Open Oobleck decoder is a convolutional autoencoder (2048× upsampling, ~21.5 Hz latent) with a receptive field of
$ \approx 8$ latent frames each side. We measure the boundary error of decoding chunk-by-chunk with `W` latent frames of context per side, versus the full one-pass decode (reference):

\begin{table}[htb]
  \caption{Lookahead Analysis in Stable Audio Open Oobleck VAE}
\label{tab:lookahead}
\small
\begin{center}
\resizebox{0.6\textwidth}{!}
{\begin{tabular}{@{}llll@{}}
    \toprule
    \multicolumn{1}{l}{\bf overlap $W$} & \multicolumn{1}{l}{\bf lookahead } &\multicolumn{1}{c}{\bf rel-L1 vs full} & \multicolumn{1}{c}{\bf peak error} \\ \midrule 
 0 (naive per-chunk) & 0 ms & 4.54\% & 0.62  \\
 4 & 182 ms & 0.39\% & 0.15  \\
 \bf 8 & \bf 364 ms & \bf 0.049\% & \bf 0.003  \\
 16 & 727 ms & 0.047\% & 0.002  \\

\bottomrule
\end{tabular}}
\end{center}
\end{table}

As shown in Table~\ref{tab:lookahead}, at $W \ge 8$ the streamed decode reproduces the full decode to the fp16 floor (peak 0.003, inaudible), because a convolution's output within its receptive field is identical once the
full context is supplied; below that, naive per-chunk decode produces audible clicks at every seam (peak 0.62). 

Therefore, our practical implementation introduces a W=8 lookahead buffer to accommodate this non-causality without breaking the overall streaming pipeline.

\paragraph{About the equivalence.} The only operator that mixes information across time in TCMamba is TimeMamba, whose SSM hidden state is a sufficient summary of the past by construction of the linear recurrence. 
The other components are pointwise in time and hence invariant to chunking: cross-attention attends only to the (fixed) text tokens, and FiLM/FFN act per frame; ChannelMamba operates within local windows of length ${sub}_l$ with no cross-window state, so it is unaffected as long as chunk boundaries align with window boundaries; and CFG is a linear combination of two such equivalent passes. 

However, exact chunk-wise equivalence requires several implementation-level alignments: window alignment, causal conv1d warmup at chunk boundaries, and consistent global time positions. We acknowledge that these cannot make the chunked and one-time paths bit-exact in practice; our implementation handles every one of these alignment points as carefully as possible, which yields an empirically very small discrepancy. We therefore only achieve "approximate equivalence with empirically negligible discrepancy".

\begin{table}[htb]
  \caption{Chunked generation v.s. one-time generation}
\label{tab:chunk}
\small
\begin{center}
\resizebox{0.8\textwidth}{!}
{\begin{tabular}{@{}llllllll@{}}
    \toprule
    \multicolumn{1}{l}{\bf Model} & \multicolumn{1}{l}{\bf Chunk-Size} &\multicolumn{1}{c}{\bf FD${\downarrow}$} &\multicolumn{1}{c}{\bf KL${\downarrow}$} &\multicolumn{1}{c}{\bf IS${\uparrow}$}   &\multicolumn{1}{c}{\bf FAD$_{v}$${\downarrow}$} &\multicolumn{1}{c}{\bf CLAP${\uparrow}$}  &\multicolumn{1}{c}{\bf FAD$_{openl3}$${\downarrow}$}
\\ \midrule 

  {\proposed}& - & 11.8569& 1.1158& 11.6865& 2.5248 & 0.4957& 96.2363 \\
  TangoFlux-chunked & $\sim$64 &21.0276 & 1.6026 & 9.6641 &3.3030 & 0.3446 & 109.1011 \\ \midrule
  {\proposed}& 32 & 12.0262 & 1.1277 & 11.3561 & 2.7999 & 0.4808& 98.0833  \\ 
{\proposed}& 64 & 11.5988 & 1.1545 & 11.8024 & 2.5961 & 0.4900& 93.6472  \\ 
  {\proposed}& 96 &  11.3551 & 1.1447 & 11.6086 & 2.5946 & 0.4890& 94.2503 \\
  {\proposed}& 128 &  11.2576 & 1.1262 & 11.5822 & 2.4817 & 0.4910 & 93.0274  \\

\bottomrule
\end{tabular}}
\end{center}
\end{table}

We conducted a quantitative analysis to validate the soundness of chunked generation. As shown in Table~\ref{tab:chunk}, the performance discrepancy between chunked generation and monolithic one-time generation is negligible.

\subsection{Out of Domain Generalization}
\label{appendix:ood}

To assess generalization, we evaluate on the Clotho evaluation set. This is a genuine out-of-domain test for our model: Clotho is sourced from FreeSound, which we do not train on, so there is no domain overlap or leakage from our side. 

\begin{table}[t]
  \caption{Performance comparison across baselines on Clotho testset.}
\label{tab:ood}
\small
\begin{center}
\resizebox{0.95\textwidth}{!}
{\begin{tabular}{@{}lllllllllll@{}}
    \toprule
\multicolumn{1}{l}{\bf Duration} & \multicolumn{1}{l}{\bf Model}  &\multicolumn{1}{c}{\bf{Parameters}} &\multicolumn{1}{c}{\bf FD${\downarrow}$} &\multicolumn{1}{c}{\bf KL${\downarrow}$} &\multicolumn{1}{c}{\bf IS${\uparrow}$}   &\multicolumn{1}{c}{\bf FAD$_{v}$${\downarrow}$} &\multicolumn{1}{c}{\bf CLAP${\uparrow}$}  &\multicolumn{1}{c}{\bf FAD$_{openl3}$${\downarrow}$}
\\ \midrule 

    \multirow{7}{*}{10s}
  & AudioLDM2-48K & 1.7B & 28.4461 & 2.4003 & 4.5949 & 5.5341 & 0.4886 & 191.7932 \\
& Stable Audio Open & 1B & 29.7914 & 2.3376 & 8.4862 & 2.3854 & 0.4895 & 86.9946 \\
  & TangoFlux-base &  516M & 20.9078 & 2.210 & 8.7841 & 2.8864 & 0.4796 & 85.5309 \\ 
  &\cellcolor{highlightc!60}{\proposed}-base & \cellcolor{highlightc!60}581M& \cellcolor{highlightc!60}\first{16.7074} & \cellcolor{highlightc!60} \first{2.1671} & \cellcolor{highlightc!60}8.7764 & \cellcolor{highlightc!60} \first{2.1386} & \cellcolor{highlightc!60}0.5031 & \cellcolor{highlightc!60} \first{73.8142} \\ 
  &TangoFlux &  516M & 21.0051 & 2.2747 & \first{9.0996} & 2.4150 & 0.5014 & 78.1923 \\
  & \cellcolor{highlightc!60}{\proposed} & \cellcolor{highlightc!60}581M & \cellcolor{highlightc!60}19.6883 & \cellcolor{highlightc!60}2.2069 & \cellcolor{highlightc!60}8.7865 & \cellcolor{highlightc!60}3.1331 & \cellcolor{highlightc!60}\first{0.5058} & \cellcolor{highlightc!60}82.2531\\

  \bottomrule
\end{tabular}}
\end{center}
\vspace{-5mm}
\end{table}

As shown in Table~\ref{tab:ood}, MFlowAudio generalizes well to this different-domain distribution, outperforms other baselines.

\subsection{Quantitative Evaluation of Transition Smoothness}
\label{appendix:transition}

Audio is inherently a time-continuous signal; consequently, temporal discontinuities are highly perceptible to the human auditory system. Naive concatenation of two independently generated audio chunks inevitably introduces severe acoustic artifacts, such as:

1) Phase Discontinuities (Phase Jumps): The temporal boundary of two independently sampled waveforms rarely aligns precisely at a zero-crossing. An abrupt, non-differentiable amplitude jump at the splicing point physically induces a sudden broadband energy burst. Perceptually, this translates into highly obtrusive "click" or "pop" artifacts.

2) Noise Floor Shifts: Discrepancies in ambient background noise or acoustic environments between adjacent chunks (e.g., transitioning from a chunk with a subtle hiss to a purely clean one) result in a sudden drop-off or unnatural introduction of noise. This severely breaks acoustic continuity, creating a jarring sensation akin to power being abruptly cut off.

\begin{figure}[htb]
\setlength{\abovecaptionskip}{-1pt}
\setlength{\belowcaptionskip}{-1pt}
\begin{center}
  \includegraphics[width=.61\linewidth]{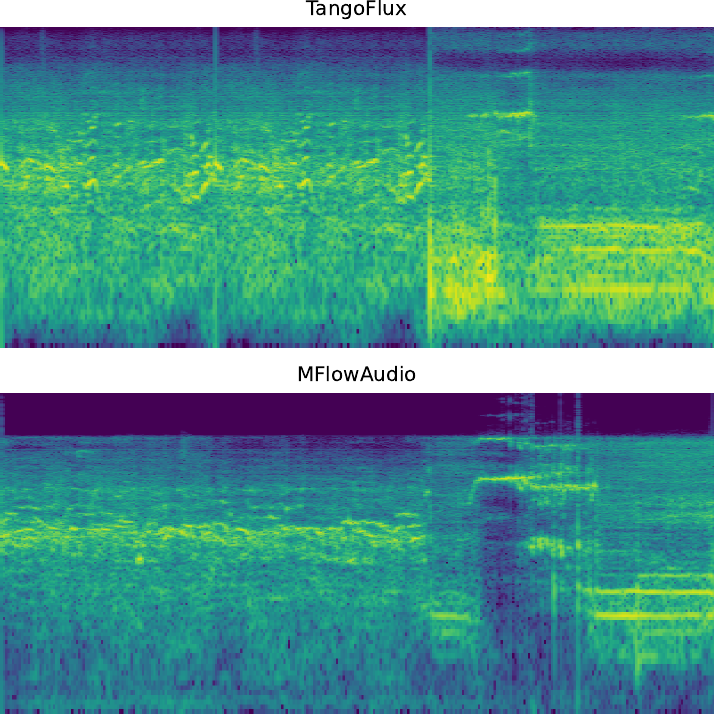}
\end{center}
\caption{Spectrum generated by TangoFlux (top) and MFlowAudio (bottom) via dynamic prompting. Prompts: "0, people are talking" "3, a crowd of people talking" "6, a horse neighing as a lawnmower engine runs in the background". There is a noticeable spectral jump in TangoFlux at the third and sixth seconds (a bright vertical line).}

\label{fig:combined}
\end{figure}

As illustrated in Figure~\ref{fig:combined}, the top spectrogram presents pronounced bright vertical striations at the temporal boundaries. These striations indicate severe structural fragmentation and abrupt spectral discontinuities across the transition.

To quantify this unnaturalness during dynamic prompt switching, we formulate three targeted evaluation metrics based on the concept of spectral flux. Collectively defining this framework as Boundary Spectral Flux, the formulation is detailed as follows:

\begin{definition}[Boundary Spectral Flux Metrics]
Let $\mathrm{SF}^{(i)}[m]=\sum_{\omega}\max\!\big(X^{(i)}_{m+1}[\omega]-X^{(i)}_m[\omega],0\big)$ be the half-wave-rectified spectral flux at frame $m$ of sample $i$ (STFT with Hann window, $N{=}2048$, hop $H{=}512$), and $\tau_m = (m+1)H/\mathrm{sr}$. Given prompt-switch times $\{t_k^{(i)}\}_{k=1}^{K_i}$ and window half-width $\Delta$, the three reported quantities are defined as follows:


\begin{equation}
\begin{aligned}
\mathrm{BMaxSF} &= \tfrac{1}{N}\!\sum_{i=1}^{N}\tfrac{1}{K_i}\!\sum_{k=1}^{K_i} \max_{m:\,|\tau_m-t_k^{(i)}|\le\Delta}\mathrm{SF}^{(i)}[m], & \text{(boundary peak)},\\
\mathrm{BgMedSF} &= \tfrac{1}{N}\!\sum_{i=1}^{N}\mathrm{median}\!\big\{\mathrm{SF}^{(i)}[m]:m\in \big\{m:\,|\tau_m-t_k^{(i)}|>\Delta,\;\forall k\big\}\big\}. & \text{(background frames)}\\
\mathrm{Ratio}  &= \mathrm{BMaxSF}\,/\,\mathrm{BgMedSF}
\end{aligned}
\end{equation}

\end{definition}

\subsection{Parameter Analysis of State Decoupled Conditional Guidance}
\label{appendix:sdcg}

As articulated in Proposition~\ref{prop:vfi}, guaranteeing acoustic continuity in chunk-wise generation inherently necessitates strict consistency of the vector field at each integration step. However, when the semantic condition (i.e., the text prompt) transitions dynamically between adjacent chunks, this fundamental continuity is inevitably disrupted.
The cached state exerts historical inertia, forcefully constraining the generation trajectory to converge toward a subset of the manifold that is acoustically coherent with the preceding chunk, while the newly injected prompt attempts to redirect the trajectory's endpoint toward a different semantic target space. Consequently, vector field collision precipitates at every single step of the ODE integration process.

To circumvent this antagonistic interference and successfully steer the trajectory toward a highly specific, unified manifold that optimally satisfies both temporal continuity and the newly introduced semantic constraints, we employ SDCG to dynamically calibrate and orthogonalize the target vector field at each integration step.
Since the historical state might go against the guidance of the new prompt in scenarios of semantic mismatch, we introduce a scaling factor, $\delta$ (as formulated in Equation~\ref{eq:sdcg}), to explicitly modulate the flux of the historical state.

\begin{table}[htb]
  \caption{Performance varies with $\delta$}
\label{tab:params}
\small
\begin{center}
\resizebox{0.93\textwidth}{!}
{\begin{tabular}{@{}lllllllllll@{}}
    \toprule
    \multicolumn{1}{l}{\bf $\delta$} &\multicolumn{1}{c}{\bf FD${\downarrow}$} &\multicolumn{1}{c}{\bf KL${\downarrow}$} &\multicolumn{1}{c}{\bf IS${\uparrow}$}   &\multicolumn{1}{c}{\bf FAD$_{v}$${\downarrow}$} &\multicolumn{1}{c}{\bf CLAP${\uparrow}$}  &\multicolumn{1}{c}{\bf FAD$_{openl3}$${\downarrow}$} & \multicolumn{1}{l}{\bf BMaxSF}  &\multicolumn{1}{c}{\bf{BgMedSF}} &\multicolumn{1}{c}{\bf{BBR}${\downarrow}$}
\\ \midrule 

1.0 &16.0462& 1.2453& 7.6155 & 2.9257& 0.5071& 102.2526& 852.1209 & 199.0192 & 4.2816 \\
0.9 &16.0299& 1.2422& 7.6105 & 2.9177& 0.5072& 102.1618& 852.9687 & 199.1520& 4.2830 \\
0.8 &16.0364& 1.2387& 7.6059 & 2.9077& 0.5074& 102.0876& 854.3832 & 199.2957& 4.2870 \\
0.5 &15.9700& 1.2310& 7.6002 & 2.8672& 0.5078& 101.7815& 857.0271 & 199.6675& 4.2923 \\
0.1& 15.8575& 1.2190& 7.5717 & 2.8448& 0.5087 & 101.2825 & 865.8578 & 200.0253 & 4.3287 \\

\bottomrule
\end{tabular}}
\end{center}
\end{table}

The empirical parameter analysis is presented in Table~\ref{tab:params}. Given that our evaluation prompts were not explicitly engineered for sequential coherence, inherent semantic conflicts across boundaries are pronounced. Consequently, decreasing $\delta$ (thereby attenuating the state's inductive bias) significantly enhances semantic alignment and distributional fidelity. However, this relaxation entails a corresponding decline in the Boundary-to-Background Ratio (BBR). This degradation confirms that overly penalizing the state flux compromises temporal continuity, precipitating unnatural acoustic transitions. To achieve an optimal trade-off between semantic responsiveness and boundary smoothness, thoughtfully engineered prompt trajectories are necessary.

\subsection{The Efficacy of ChannelMamba}

To evaluate the contribution of the ChannelMamba module, we conducted an ablation study by removing it entirely from the TCMamba architecture (denoted as the No-Channel baseline). Notably, this specific module is lightweight, introducing a marginal parameter overhead of merely ~2M (e.g., increasing the total parameter count from 581M to 579M).

\begin{figure}[htb]
\begin{center}
  \includegraphics[width=.45\linewidth]{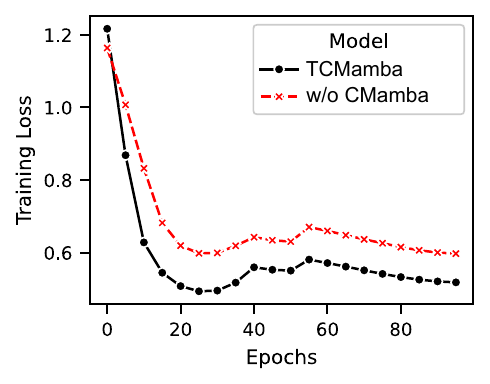}
\end{center}
\caption{The loss curve during training.}
\label{fig:train}
\end{figure}

Despite its minimal parameter footprint, its impact on model performance is significant. As illustrated by the training loss trajectories in Figure~\ref{fig:train}, the No-Channel configuration exhibits substantially slower convergence and ultimately plateaus at a higher loss floor compared to the full model equipped with ChannelMamba.

\begin{table}[htb]
  \caption{Ablation for ChannelMamba}
  \label{tab:abl-freq}
\small
\begin{center}
  \begin{tabular}{@{}llllllll@{}}
    \toprule
\multicolumn{1}{l}{\bf Duration} & \multicolumn{1}{l}{\bf Model}  &\multicolumn{1}{c}{\bf{Parameters${\downarrow}$}} &\multicolumn{1}{c}{\bf CLAP${\uparrow}$} &\multicolumn{1}{c}{\bf IS${\uparrow}$} &\multicolumn{1}{c}{\bf FAD${\downarrow}$} &\multicolumn{1}{c}{\bf FD${\downarrow}$} &\multicolumn{1}{c}{\bf KL${\downarrow}$}
\\ \midrule 

\multirow{2}{*}{10s} &  w/o ChannelMamba    & 579M  & 0.3758 & 10.0166 & 6.9879 & 18.1077 & 1.1983  \\
                     & {\proposed}         & 581M & 0.4655 & 11.6865 & 2.5248 & 11.8569 & 1.1158 \\

  \midrule

\multirow{2}{*}{30s} &  w/o ChannelMamba    & 579M  & 0.2596 & 5.5623  & 9.0628 & 45.6336 & 1.3096 \\
                     & {\proposed}         & 581M &  0.4448 & 11.3356 & 2.8981 & 17.5387 & 1.1411 \\

\bottomrule
\end{tabular}
\end{center}
\vspace{-2mm}
\end{table}

Empirical evaluations on the test set further corroborate this observation. 
As shown in Table~\ref{tab:abl-freq}, the No-Channel model suffers a severe degradation in both semantic alignment and acoustic generation fidelity. Crucially, this performance gap widens drastically during sequence length extrapolation (e.g., 30 seconds of audio), revealing that spectral modeling is indispensable for maintaining long-term structural coherence. 
In sum, these findings underscore that the lightweight ChannelMamba module yields massive gains in both training stability and generative capacity.

\end{document}